# THE EFFECTS OF TECHNOLOGY DRIVEN INFORMATION CATEGORIES ON PERFORMANCE IN ELECTRONIC TRADING MARKETS.

**JIM SAMUEL**
Jim@aiKnowledgeCenter.com

**RICHARD HOLOWCZAK**
Richard.Holowczak@baruch.cuny.edu

**ALEX PELAEZ**
Alexander.Pelaez@hofstra.edu

## ABSTRACT

Electronic trading markets have evolved rapidly with continued adoption of new technologies and growing information acquisition and processing capabilities. Traditional perspectives on trading performance adopted a monolithic view of information. Past research and practitioner heuristics posit that adopting new technologies and incorporating more information should increase price efficiency and trading performance uniformity. However, along with technological change, information dynamics have evolved significantly resulting in immense growth in data volumes, and increased complexity of information categories. The present research explores behavioral trading performance under varying information category conditions and argues that unfettered technological developments and information consumption will not necessarily lead to consistent improvement in uniformity of trading performance.

In this study, we employ an artificial stock market based economic experiment to examine the role of technology driven information categories in influencing trading decisions in electronic markets. Financial electronic markets are used as an information-rich mature markets representation to analyze information category driven trading performance. The results show that a variation of information categories can influence trading performance. The findings provide a basis to better understand behavioral phenomena in electronic markets and can be used to explain anomalies as well as to manage trading performance in electronic markets.

## INTRODUCTION

*"This is the information systems predicament*
*- using information as a ubiquitous label whose meaning is almost never specified."*
-       McKinney & Yoos (2010)[36]

'Information' has been recognized as a critical factor in research across domains, business strategies, markets and social institutions. However, it is evident that the idea of information, its characteristics and tangible artifacts that represent information vary widely across research and practitioner domains. Systematic efforts to study the effects of such variations in information characteristics and tangible information artifacts have been limited. Electronic trading markets (ETMs) constitute one of the key information sensitive domains, where information has been subject to a monolithic perspective in most of the early and popular research. Information acquisition and information processing have become mostly technology driven in present markets. Technology driven information is therefore a critical dimension of present day electronic markets. It is important to continually improve our understanding of how technological advances and information dynamics influence human decision making and electronic trading performance. The present research seeks to provide clarity on the technology driven information dimension of ETMs by studying if variation in information characteristics and tangible information artifacts have any effects on human decision making and trading performance.

Over the last several decades, there has been a dramatic increase in the way technology is used in ETMs with the intent to increase operational efficiency and gain competitive advantage. Many market participants are caught in a technological and informational arms race, believing that the continued adoption of new technologies and more information will lead to competitive advantages and thus result in performance gains. On a market level, the underlying heuristic has been that more information will increase price efficiency in ETMs, without sufficient mention of the potential role of information facets and categories [25] in influencing price efficiency and performance in electronic trading markets. Here 'performance' is used as a comparative objective measure of periodical monetary gains through trading, in the context of the performance of comparable traders. Behavioral trading performance (BTP) is an important component of ETMs [8] and generally refers to variations in trading performance of participants in ETMs based on the influence of behavioral factors, varying beliefs, personalized infor­mation processing, technological and market environment factors, and bias [9][1]. While trading performance is an objective monetary measure, BTP identifies the performance factors as being behavioral in nature.

Advances in information theory and perception challenge both the absence of attention to information characteristics as well as the simplistic equalization of the effects of various information categories [19]. 'Information categories' refer to domain specific information classification, such as the classification of information as being 'public' or 'private' information in the capital markets domain. Grover et al. [25] have effectively used the term "facet" to classify categories of information artifacts. In spite of an absence of clear contrast between the two terms, there is a broader use of the term "category" [38] to classify information artifacts in information management studies and the present research thus uses the term "information categories" to classify information artifacts based on ETM domain specific characteristics. 'Technology Driven Information Categories' (TDIC) is used to reflect the ontological structure of information artifacts (tangible expressions of information such as news websites, live streaming of prices or software used to create information by standard or proprietary data analysis), the creation, storage, communication, management and analysis of which are largely driven by technology in ETMs. It must be noted that 'information category' is ontologically different from 'information format'. 'Information format' refers primarily to the presentation characteristics of information (i.e., 2D versus 3D line graphs in Kumar & Benbasat, [32]) unlike 'information category' which is primarily associated with characteristics such as the source of the information, timeliness and beliefs gen­erated based on the degree of exclusivity or reliability of informational content in the information artifacts.

Information has been accepted to be a key driver of price transparency in electronic markets [25]. Market efficiency has been defined by the degree of ability of a market to facilitate prices which fairly, fully and in real time, represent all available information [17]. A pervasive belief amongst practitioners and researchers is that information improves market effi­ciency and that markets rapidly reflect information in corresponding asset prices [50][18]. Thus by definition, market efficiency is built on informational influence with a traditionally monolithic view of information, and the effects of varying information categories have been mostly ignored in early studies on market efficiency. An increased awareness of the significance of infor­mational influence has led to significant investments into technology and information management in global markets over the past few decades. Therefore, if the efficient markets hypothesis (EMH) were fully representative of market efficiency dynamics, then the tremendous increase in operational efficiencies through technological advances in markets along with historically unparalleled high information acquisition management levels should have led to near optimal market efficiency as would be evidenced by steady security prices and fair uniformity of performance in ETMs (i.e., there would not be any significant trading performance variation amongst market participants). However, studies in intraday price volatility in equity markets indicate varying volatility patterns [40]. Similarly, studies covering longer time intervals over the past decade [15], do not show any significant reduction in market volatility during a period which has seen significant adoption of technological advances and strong growth in information management in ETMs. This along with observed variation in the annualized trading performance of participants in ETMs challenges classical beliefs regarding the role of information in markets. Researchers have attempted to address sustained price volatility and variance in trading performance by revisiting traditional efficiency theories and pro­posing new frameworks, including behavioral and hybrid explanations [33].

The present research takes into account recent developments in information theory [19] and information systems [25], and studies the influence that TDIC can have on performance in ETMs. Electronic equity markets represent one of the most mature and sophisticated electronic market environments, and are therefore considered as being fairly representative of the general characteristics of ETMs at large. This is especially so from an information based pricing perspective, taking into account the large volumes of information and information processing capabilities that are becoming increasingly available to common ETM participants. The study is designed to provide experimental evaluation of TDIC influencing performance in ETMs as this can have meaningful and actionable implications for furthering our understanding of price movements and BTP.

The term "Information categories" is generally used to classify information artifacts with distinctive properties into groups. We define four main information categories under ETM domain specific TDIC: 1) price information, 2) public information, 3) private information (distinguished from insider information) and 4) privatized (generated by ETM participant using interactive IT tool) information, each of which are elaborated upon in the description of the experiment. We employ an economic experiment, using a stock market simulation [51] to study the effects of technology driven information categories (TDIC) upon behavioral trading performance (BTP) in electronic trading markets (ETMs). This work builds upon research from multiple domains and therefore references literature in information systems, equity markets, electronic commerce and information science domains.

# LITERATURE REVIEW

Information Systems (IS) theories, notably the EMH (Electronic Markets Hypothesis) [34] posit that technology is expected to make markets more efficient by reducing coordination costs and increasing price transparency [21][22]. IS and related research also points to technology reducing search costs [56][57] and enabling advanced communication [12]. Given the voracious market appetite for new technologies and more information, markets should demonstrate increasing informational efficiency: this should logically be reflected in relatively and steadily increasing stability of prices (lower volatility) and improved pricing mechanisms [41][43][2]. Advances in electronic trading technologies and information availability are therefore expected to increase informational efficiency, support price transparency and thus lead to trading performance uniformity. However, evidence points to the contrary [15]. While volatility and variances in trading performance may be attributable to various economic and external factors, the present research posits, by controlling other factors using a trading experiment, that technology driven information categories can also contribute to variance in trading performance. Researcher and practitioner perspectives on markets which are based on traditional definitions of information (which do not account for the influence of varying technology driven information categories), cannot be used to understand ETM phenomena which arise in a significantly advanced information ecosystem as we have today [25]. Therefore there has been a steady increase in research focus on the influence of information characteristics.

Elaborating upon the TDIC, price information is representative of the knowledge of current prices of specific securities or stock. This could be obtained on a trading floor or by other means such as access to a live trading platform. Historically, various news channels provided information to investors and traders, and this information formed the basis for trading decisions. 'Public information' is representative of this and other publicly available information sources. 'Private information' is representative of proprietary analysis and associated research reports which some traders may have access to. Generally, traders expect superior performance for trading based on such private information.

Large investments have been made into developing and deploying technologies, that help market agents - traders, investors and market participants process commonly available information artifacts, using uniquely designed information analysis tools with varying degrees of sophistication. These tools enable a personalization of agent expectations of price movement. The fourth TDIC 'privatized information' is representative of such proprietary tools which allow traders to discover and thus 'privatize' information based on their customized usage of technology. There has thus been a move from a historically monolithic view of information which was mainly available publicly as news and numbers, to the present time, where we see a high degree of technology enhanced information acquisition, processing and personalization that has the potential to variably affect decision-making and trading performance in ETMs.

Heuristically, it is clear that variance in information content may lead to variance in performance (superior information can enable a trader to gain superior returns). However, here we hold the information content as a constant across all four TDIC. This means that if the information content of the first TDIC (price only) indicates a starting price of $10 and upward price trend, then the exact same starting price of $10 and upward directionality will be implied by all the other 3 TDIC. Only the information categories (the four TDIC) are varied, to convey the same information content (such as starting stock price $10), to impact trader perception and analyze corresponding BTP. In this manner, the artificial stock market is used to simulate trading experiences to explore the role of TDIC upon trading performance based on the trader's perception of the information

category, which helps us to address a critical question: Do technology driven information categories (TDIC) influence behavioral trading performance (BTP) in electronic trading markets (ETMs)?

TDIC's influence on BTP would be evidenced by a measurable objective change in trading performance operationalized as monetary gains. If a variation in TDIC does influence performance in ETMs, then the management of information categories becomes significant for electronic trading across domains. A trading experiment in an artificial stock market is used to address the above question. An improved understanding of information category influenced performance can help us better understand behavioral factors in price movements and trading performance in ETMs, especially in markets with a higher degree of sensitivity to information and behavioral factors. We develop conditions to simulate four TDIC conditions: price-only, public, private and privatized, which can be grouped into two higher level information categories: 'price only' and 'public' information conditions are both public information conditions, while 'private' and 'privatized' are both private information conditions. Ultimately, we seek to experimentally explore if trading performance is challenged by information categories, specifically domain specific TDIC, then such information categories can lead to price movements - a 'price dance', even in the absence of fundamental changes in markets. The resulting price fluctuation would be reflected in variations in BTP by market participants under varying information categories. This will help address the question: Is such a dance a predictable Foxtrot or an unstructured and random street dance? If predictable, the TDIC will result in systematic performance effects, if random then we may not be able to observe TDIC effects.

# HYPOTHESES DEVELOPMENT

Grover et al. [25] address the "dark side of information" where the authors posit that variance of information characteristics leads to variance in prices and valuation perspectives. This is a shift from classical positions on the influence of information as seen in Hayek [16], who posited that market prices will reflect all available information in the market, though individual level information may be unique, using 'information' monolithically. Fama [17] posited informational efficiency of markets in the "efficient markets hypothesis", again using 'information' monolithically, and these studies suggest the sufficiency of price as a fair information driven indicator of asset value. If so, there should be no difference in trading performance or price dispersion based on price only information, public information and private information conditions. However, investors tend to under-react to information artifacts which they believe to constitute public information and over-react to information artifacts which they believe to constitute private information [27].

Studies in finance and market microstructure [11] address the issue of divergent expectations where even in mature markets, investors may interpret information differently based on uniqueness of perspectives. Market participants can interpret the same information differently based on their beliefs about the source of information and also process information using unique and proprietary tools and valuation models which can leads to differences in fair value estimates of asset prices. Such 'discrepancies' can occur in spite of the same information being available at the same time to market participants and thus tends to be aligned with the main concepts of the challenges faced in interpreting information categories [25]. Other studies [6] also indicate such discrepancies in supposedly mature electronic markets. Traders and investors form beliefs and preferences based on information. Even given the same objective information, traders and investors can display different trading choices and behavior [44] based on differing beliefs and expectations. Past research has highlighted various behavioral effects including the tendency to be unjustifiably optimistic [54] and poor associative-representative judgment [30]. Prior IS research has also pointed to user perspectives being shaped by information categories in spite of essential content remaining the same [23]. The discussion above leads us to posit that differences in information categories can affect the perceptions and decisions of market participants and also that ETM participants respond to varying information sources differently, hence we hypothesize that:

**H1:** *Public information based trading will result in higher trading performance levels than trading based only on observation of stock price movements.*

There are differences in investor perception of, and response to, private information (not the same as insider information) as compared to investor perception of and response to public information. We define 'Private information' as the information derived from proprietary analysis, information modeling and expert opinion (excluding insider information and all legally disallowed information) based on all legitimate information artifacts, which are often unique to distinct market participants. There have been a number of important studies on the role of information in equity markets [24][37][14]. These studies have demonstrated the importance of information in markets and distinguished between informed and uninformed trading [50]. It is fairly well established that new and meaningful information can be used to generate superior trading profits. Coval [10] addressed information asymmetry to show how investors trading with differential information led to price equilibrium. Different perspectives have been developed: public information benefits all participants [13], informed traders profit at the expense of uninformed traders [13], possibility that both uninformed and informed traders benefit eventually [53], and that too much information can lead to price discrepancies [26][29]. Easley-Ohara [14] specifically compared the role of 'information structure', contrasting public and private information to evaluate impact on a firm's cost of capital.

While public information is commonly understood to consist of announcements and news, private information has been used in past research in a variety of ways. Tookes [50] posits that private information could be information possessed by "sophisticated traders", distinct from insider or privileged information held by corporate "insiders". Stanislav [47] also identifies "expert networks" with specialized expertise based information, which is also distinct from insider information. Therefore, private information, as a distinct information category, can be defined as 'information' consisting of knowledge based on customized information acquisition, proprietary analysis, information modeling and expert opinion. Thus, private information can often be unique to distinct market participants and can, by definition, present the possibility to provide superior trading performance [42][5][50][9]. We focus on the behavioral aspect of trading as we expect participants to have greater belief in the value of private information and hence trade with greater confidence when using information artifacts they believe to constitute private information, while under-reacting to information artifacts they believe to constitute public information [27], even though the informational content and implications are held constant across the public and private information conditions. Hence the second and third hypotheses for this study:

**H2:** *Private information based trading will result in higher trading performance levels than public information based trading.*
    and
**H3:** *Private information based trading will result in higher trading performance levels than price-only information based trading.*

Performance improvements have been demonstrated based on the type of technology used to address a task [20]. Past research has also extended the Vessey and Galletta [52] cognitive fit model to contrast the effects of 'list' with 'matrix' type information artifacts [28]. Our focus for the final hypothesis is on studying and comparing the effects of private and privatized information categories on trading performance in ETMs by contrasting passive and interactive information categories. Hence we further qualify the understanding of private information to define 'privatized information' in this study, as being a distinct user generated information category, consisting of user-personalized or user-customized knowledge based on the usage of proprietary tools and information modeling. Expert opinion, formulation and specialized knowledge along with customization options is expected to inherent in the tools available to the user. Though we distinguish between 'private' (received-private information) and privatized (self-generated-private information), we maintain that both these categories are conceptually private. Understanding the effects of these two TDIC help address an important trading performance question from an information categories perspective: Given an ETM environment, how do interactive (such as a software program) and passive (such as a plain text report) information artifacts affect BTP?

Numerous studies have indicated information format and presentation based effects – specifically interactive (human data entry or selection and artifact management generates output) decision aids and information artifacts have been associated with improved decision performance [39][55]. Past research has also noted improved satisfaction with the use of dynamic interactive blink mechanisms [31] and decreased performance in interactive decision support systems with relatively fewer features [48]. Technology driven interactivity has also been shown to improve learning performance [3][4]. However, the role of interactive information artifacts has not been studied within the context of ETMs. This presents an interesting opportunity to contrast the effects of self-generated (interactive) privatized information and received (passive or non-interactive) private information categories. Therefore, we compare BTP based on a non-interactive information artifact (such as a private information report) with the BTP based on an interactive information artifact (such as a participant managed proprietary valuation model). Thus, from an information category analysis perspective, we classify the broader 'private information' concept into two distinct information categories: The first is the 'private information category' in the form of a proprietary report with plain text content. Technology is used to present such a report (in the form of a webpage) but there is no usage of or active participant involvement with the technology to generate or customize results. The effect of this category of private information on electronic trading performance is hypothesized below. The second is what we classify as 'privatized information category' in the form of a proprietary valuation model, requiring active participation and engagement of the user. In this case, the technology artifact requires active human management and interaction with the technology to customize and generate the results. Past research has affirmed that the use of an interactive information artifact leads to better domain level performance [49]. We expect participants to have greater belief in privatized information due to their direct involvement in generating customized information and hence trade with greater confidence and engagement when using privatized information artifacts, while having relatively lower engagement when trading based on information artifacts they believe to constitute private information Therefore, we posit that ETM participants who make use of an interactive privatized information artifact will have superior trading performance. This is hypothesized as:

**H4:** *Interactive information based trading will result in higher trading performance levels than non- interactive information based trading.*

Recent studies have demonstrated that 'information frequency' (rate of arrival of information) can affect trading performance [35] without addressing the role of information categories. The present study creates a better understanding of the role of information categories in electronic markets, thereby drawing attention to the need to qualify the category of information artifacts used in past research. Also, the use of an artificial stock market based experiment is expected to support arguments in favor of the behavioral (role of beliefs about information categories) approach to performance in electronic equity markets and ETMs at large. Furthermore, this study uses the experimental framework to control external and environmental influence, and is thus able to clearly focus on the BTP effects of TDIC in ETMs. This justifies the specific elaboration of information categories as shown in TDIC in equity markets, which is used to qualify the classical "Price information alone is sufficient" position.

# METHODOLOGY

In line with prior research using laboratory experiments to understand market dynamics, we follow the principles of experimental economics [45] by conducting a laboratory experiment using an equity market trading simulation. Equity markets were simulated with each trader (subject) in an independent artificial stock market simulation. In accordance with the induced value theory of experimental economics [46], participants were economically motivated with cash rewards based on trading performance so as to induce realistic efforts in performance. The experiment focused on exploring the influence of TDIC upon trading performance in ETMs. TDIC were presented to participants using a custom designed website and TraderEx, an electronically simulated equity trading platform [51]. Participants were allowed to trade one stock in the equity market simulation under varying TDIC. The studies were conducted on a between-participants basis.

It must be noted that within this experiment, the information content and implications of information artifacts in the TDIC conditions are not varied across conditions, except as is necessarily required to manipulate the informational conditions. 'Informational content and implications' refer to the meaning and objective knowledge contained in the artifacts used to represent the various information categories. Informational content is held constant across conditions (such as the information about the stock price being held constant at "$10" across all TDIC, though one artifact may present the information as public news and another as a private report). This is necessary because any significant change in fundamental information content (not the presentation category) could by itself result in performance changes and confound the results. However, since the informational content and implications are held constant, any or all changes in performance levels can be fairly attributed to being effect of corresponding information category conditions implemented in the experiment. BTP across TDIC is captured by the objective monetary trading performance measure which is calculated by using an end of trading day summation of profits (or losses) from trades conducted during the simulated trading day. It is expected that trading profits should be greater for the private information categories as compared to the public information categories.

## Experimental Design and Procedure

In the experiment, we implemented four information category conditions as described in TDIC: trading-system (price-only) information through the interactive TraderEx trading simulation platform, public information in the form of news on webpage custom designed for the experiment and privatized information in the form of a downloadable customized report from the same website. The fourth information category, "privatized information" was implemented as an interactive valuation model, downloadable from the experiment website. Though one common website was used for the experiment, participant access was restricted to the page containing the information artifact relevant to the one specific TDIC they were assigned to. Manipulation checks were conducted for the experimental artifacts and the checks validated the difference in perception of the information artifacts used to represent the TDIC for the experiment. The first TDIC, price-only condition provided the participants with access to trading platform only, with no additional information artifact. In the subsequent conditions, information artifacts were introduced separately for each condition, on a between participants basis.

Participants in the experiment were undergraduate students from a large, urban public university. There were 116 participants who were recruited through the school's research subject enrollment system. After registration, the participants were presented with the experimental details upon their arrival at the laboratory where the experiment was to be conducted. Participants were given a demonstration of the trading software platform "TraderEx". The trading mechanism, price information, buy and sell functions of the TraderEx platform were introduced to the participants. The trading-day-profit performance measure was explained along with a demonstration of how performance was calculated and questions raised by the participants were addressed. The live trading demonstration was followed by two practice rounds where the participants had an opportunity to familiarize themselves with the use of TraderEx. Participants were then introduced to trading conditions on a between participants basis.

The first condition was the price only condition in which participants traded only on the basis of the price information visible in TraderEx and no additional information artifact was provided. In the second information condition "public information" was presented to the participants. Participants were directed to a custom designed website with a public news report page and it was announced that this information was available to all market participants. In the third condition, "private information" was presented to the participants. Participants were instructed to access an information artifact (downloadable report) and it was announced that this information report was available as a uniquely privatized report for each subject. Participants were thus led to believe that their privatized information report was a unique analysis, in the sense of the same information not being available to other market participants. In the fourth and final condition, an interactive private information tool was made available to the participants. Participants were instructed to access an information artifact (downloadable interactive valuation model) and the simple valuation artifact's usage was demonstrated. Participants had to make a simple entry of the starting price into the valuation model to see the output. It was announced that this interactive valuation model was custom designed for the participants, to provide a fair valuation of the equity to be traded. Participants were thus led to believe that the interactive valuation model provided accurate and unique output, in the sense of the same output information not being available to other market participants. The experimental conditions are summarized in Table 1 below.

Table 1: Experimental Conditions for Information Categories

| Condition | Information Category | Description |
|---|---|---|
| c1 | Price only info | Trading platform – No additional information artifact |
| c2 | Price info + Public info | Trading platform + Website (Public News) |
| c3 | Price info + Private info report | Trading platform + Private Information – Custom Report |
| c4 | Price info + Private info tool | Trading platform + Private Information –Interactive Model |

The price-only condition, based on participants using the trading screen and observing prices, served as the base condition with no additional information artifacts and remained constant through the experiment. In three subsequent TDIC conditions, specific information artifacts were introduced and the trading performance was measured for all four conditions. This structure is aligned with the way information is received in and managed in real world ETMs, where traders have continual access to their trading screens (and are thus perpetually exposed to real-time prices) as they process and leverage additional information. The conditions were implemented in different sessions (between participants framework) and participants were tasked to trade with the objective of maximizing realized profits by the end of the simulated trading day. Trading performance was recorded for all participants at the end of the trading session and compared to identify the top performers. In line with the principles used in economic experiments [46], a cash payout ($5, $10 or $50 based on performance) was announced for the top

performers to induce appropriate motivation to trade for economic benefits. The experiment was funded by a research award from the public university.

Data from the experiments were downloaded from TraderEx at the end of each experimental session for trading performance evaluation and data analysis. The four TDIC conditions served as the independent variables in the experimental simulation: price only, public news, private information (report) and private information (interactive model). Trading performance, as the dependent variable in our analysis, was measured based on individual end-of-trading-day profits. Participants were expected to buy and sell units of a single stock in a simulation of an electronic market over a trading day. The trading day was simulated in 10 minutes (accelerated mode) of trading and participants were expected to buy and sell, starting and ending the day with a net quantity of zero units, during the trading day based on the information condition they were assigned to. A one-way ANOVA was used for the primary analysis consisting of four levels for information categories.

## ANALYSIS AND RESULTS

A total of 116 undergraduate students from a large, public urban university participated in the experiment. Due to the nature of the subject pool and the rules governing the use of the pool, complete anonymity was maintained to support ANOVA conditions requirements. Table 2 summarizes the descriptive analysis of the experimental data for each of the conditions. A negative mean implies poor trading performance and a positive mean implies relatively better trading performance. Conditions c1 and c2 (price and public information conditions respectively) have high standard deviations for performance measure indicating that participants in these conditions perceived the market with far less homogeneity than participants in the c3 & c4 (passive and interactive privatized information) conditions.

Table 2: Information Categories - Summary of Descriptive Statistics

|  | c1 | c2 | c3 | c4 |
|---|---|---|---|---|
| Mean | -229.793 | -162.004 | 7.020707 | 31.34142 |
| Standard Deviation | 308.7807 | 156.3064 | 26.99493 | 61.43807 |
| Median | -85.7999 | -93 | 3.90003 | 29.60031 |
| Subject Count | 27 | 28 | 29 | 29 |

The study hypothesizes a significant improvement in BTP for information conditions where the trading performance based on public information is expected to be better than trading performance based on price-only information, the trading performance based on private information is expected to be better than trading performance based on public information and the trading performance based on privatized information is expected to be better than trading performance based on private information. Public, private and interactive information conditions are all thus hypothesized to have better trading performance than the price only condition. The hypotheses can represented by the following equations:

$$\alpha[pu] > \alpha[po] \qquad \ldots(1)$$
$$\alpha[pv] > \alpha[pu] \qquad \ldots(2)$$
$$\alpha[pv] > \alpha[po] \qquad \ldots(3)$$
$$\alpha[pvd] > \alpha[pv] \qquad \ldots(4)$$

…where, $\alpha[po]$ represents the average trading performance in condition "c1", the price-only information category; $\alpha[pu]$ represents the average trading performance in condition "c2", the public information category; $\alpha[pv]$ represents the average trading performance in condition "c3", the private information category and $\alpha[pvd]$ represents the average trading performance in condition "c4", the privatized information category. Trading performance across these paired conditions are best tested by analyzing and contrasting the average performance across the conditions. The first step in this process is to test for differences in means of the trading performance effects from the four conditions. An ANOVA was used to study the differences in trading performance averages across TDIC (conceptually, the significance of $\alpha[pvd] \neq \alpha[pv] \neq \alpha[pu] \neq \alpha[po]$). Subsequently we further explore an integrated perspective of the hypotheses which can be expressed as an equation using inductive logic on equations 1, 2, 3 and 4 above, leading to the following equation:

$$\alpha[pvd] > \alpha[pv] > \alpha[pu] > \alpha[po] \qquad \ldots(5)$$

An ANOVA was used to compare performance, based on the four experimental conditions. The primary ANOVA analysis shows that the difference in conditions is significant (F=15.27, p<.01) as shown in Table 3.

Table 3: Information Categories ANOVA - Variance Analysis

|  | Df | Sum Sq | Mean Sq | F value | Pr(>F) |
|---|---|---|---|---|---|
| Group | 3 | 1362934 | 454311 | 15.27 | < 0.0002 |
| Residuals | 110 | 3272640 | 29751 |  |  |

As posited in the theoretical arguments leading to the development of the hypotheses, our analysis finds significant differences between trading performances in the four TDIC conditions. The ANOVA tested for and confirmed the significance of differences in trading performance averages across TDIC conditions, proving:

$$\alpha[pvd] \neq \alpha[pv] \neq \alpha[pu] \neq \alpha[po] \qquad (6)$$

Thus, we can conclude that within the framework of the electronic stock market experiment, the four TDIC conditions significantly influenced trading performance in the simulated ETM. This finding confirms the significance of the role of TDIC, and by itself serves as a robust anchor for further analysis and discussion. Continuing with the analysis, it is noted that the ANOVA model used to analyze the four conditions does not provide clarity on relative trading performance between individual categories. However, there is a need to analyze relative trading performance between the conditions. Since the hypotheses focus on the comparison of specific conditions, it is imperative to compare paired conditions as posited in the hypotheses and equations 1-4 above, to identify which conditions contribute to the difference in electronic trading performance. To achieve this, we employed a Tukey's HSD analysis to compare pairwise means (Table 4).

Based on the pairwise analysis results, trading performance driven by the private information category (c3) shows significant difference from the trading performance based on the price only category (c1) and so also trading performance based on the private information category (c3) shows significant difference from trading performance based on the public information category (c2). This supports the second and third hypotheses, H2 and H3. The first hypothesis (H1) states that public information based trading (c2) will result in higher trading performance levels than trading based only on observation of stock price movements (c1). However trading performance based on the public information condition (c2) was not significantly higher than the trading performance based on the price-only condition (c1) and hence the first hypothesis was not supported. This result was not entirely surprising as it does not go against the main premise for information category based trading performance presented in the research because both price-only (c1) and public information (c2) conditions are both essentially public in nature and can therefore evoke fairly aligned, though not exactly similar beliefs and trading decisions.

Unlike c1 and c2, the third and fourth information category conditions of private report (c3) and private valuation tool (c4) are both private categories and can therefore be expected to evoke dissimilar beliefs and trading decisions from c1 and c2. The second and third hypothesis state that private information based trading (c3) will result in higher trading performance levels than both, public information based trading (c2) and price-only information based trading performance (c1). The analysis supports hypothesis two and three (P value < 0.01 for each). This finding supports the main premise for TDIC based trading performance presented in the research, demonstrating that belief in private information resulted in better trading performance. This behavioral effect is further supported by comparing the trading performance in the valuation tool based privatized information category condition (c4) with the two public conditions (c1 and c2). Data analysis shows that the valuation tool based private information condition based trading (c4) resulted in significantly (P Value < 0.01) higher trading performance levels than both, public information based trading (c2) and price-only information based trading performance (c1).

The fourth hypothesis posits the superiority of interactive information based trading performance (c4) as compared to trading based on non- interactive information (c3). This position was not supported mainly due to the effect both c3 and c4 conceptually using private information (c4 is identified as 'privatized' but is still conceptually under a broader 'private information' classification). Thus the third and the fourth conditions were both conceptually private information conditions and hence it is possible that the passive –interactive contrast was superseded by the belief in the superiority of private information. This presents an opportunity for further investigation into the contextual nature of the passive –interactive information artifact effects.

Table 4: Pairwise Analysis for Information Category Based Performance

| Tukeys HSD - Info. Levels: | Diff | Lower | Upper | p adj |
|---|---|---|---|---|
| c2-c1 | 67.78904 | -53.579 | 189.157 | 0.4669019 |
| c3-c1 | 236.8133 | 116.4769 | 357.1496 | 0.0000007 |
| c4-c1 | 258.1193 | 138.7538 | 377.4847 | 0.0000008 |
| c3-c2 | 169.0242 | 49.80585 | 288.2426 | 0.0019059 |
| c4-c2 | 190.3302 | 72.09195 | 308.5685 | 0.0003149 |
| c4-c3 | 21.30602 | -95.8731 | 138.4851 | 0.9645807 |

# DISCUSSION

Given the rapid advancement of the use of technologies for acquiring, analyzing, and managing continually increasing information quantities in ETMs globally, the present research provides a unique opportunity for understanding TDIC effects in ETMs from a BTP perspective. From an equity markets domain perspective, the studies add value by providing a deeper understanding of the role of price only, public, private information and interactive privatized information categories in electronic markets and the findings of the experiment support the behavioral (role of human cognitive and psychological factors) schools of thought on trading performance in equity markets. Three key arguments justify the information systems research perspective adopted herein: first, technological dominance of market microstructure is evident and it is necessary for researchers and practitioners to gain a strong information systems perspective on the functioning of ETMs; second, 'information' as an artifact is best understood from an information systems perspective as it is the IS domain that we gain the strongest theoretical and practical insights into the integration of technology, information and human behavior; third, given that behavioral finance has already been established [44] as a significant area of research, it becomes imperative that IS theoretical frameworks are applied to study ETMs behavioral dimensions. Therefore, it is expected that the present study would stimulate additional IS research focused on studying the effects of varying information and technology artifacts on BTP.

The present research opens avenues for further research, building on Grover et al. [25], on how TDIC can impact price movements in ETMs. This research also provides IS researchers and managers a novel perspective on the role of domain specific information categories with potential implications for electronic commerce, since the study can be extended to other non-equity and non-trading electronic markets. Extended implications of TDIC can be applicable beyond ETMs to the domain of general electronic markets at large. Additional research will be required to explore TDIC effects on consumer who use electronic marketplaces to buy and consume products and services, because the we are focused on analyzing the effects of TDIC on traders who buy and sell for profit and are therefore unaffected by product brand, subjective user ratings, and similar considerations. The private report information category condition, where participants were led to believe that the information they received (though similar in content and implications as the public information condition, between participants) provided superior returns. This result does not support the strong forms of the EMH (finance, price reflect all information immediately including private and insider information). However, this result is supportive of the weak form of the EMH where only past information is reflected in market prices, allowing for profit taking with new information. Importantly, the present study validates the behavioral aspects of ETMs based on variation in TDIC: since all the participants across the four TDIC conditions received the same informational content and price implication (no objective change in the meaning of message and objective knowledge presented), therefore, trading decisions leading to superior trading performance in the private information conditions can be treated as being behavioral (no objective differences in information content were present). This justifies our claim if the present research positing that TDIC influences BTP in ETMs. The belief and perception of advantage (or the lack thereof) that was generated based on the fact that the information provided in the private information report and private information tool categories was exclusive, highlights behavioral aspects in trading performance in ETMs which need to be further detailed and researched with greater granularity.

Overall, the results are aligned with the influence of information characteristics [25] and behavioral effects in ETMs [1], demonstrating that there is a TDIC based behavioral effect influencing performance in electronic trading. Additionally, the research methodology makes an interesting contribution to the body of economic experiments literature by using multiple human participants, individually in independently simulated artificial electronic stock markets. Therefore the trading performance of one or more participants in a condition did not influence or relatively distort the markets for other participants in the same or another information condition. This ensured better control of informational effects in each condition and eliminated a significant source for confounding effects. The originality of the experiments and novel manipulation of information categories helps isolate the effect of varying TDIC conditions without changing informational content. Thus, the present research provides insights into how trading performance is influenced by information categories in electronic markets. Our extensive literature review, though not exhaustive, indicates that there has been no research initiative to date, which studied the effects of information categories in ETMs using the methods outlined herein or otherwise - this underscores the methodological uniqueness and novelty of the present research.

Three key weaknesses of the present study present opportunities for future research: the first being the limited representativeness of ETMs, the second is the limited number of subjects, groups and experimental conditions and the third is the use of a fairly homogenous student participant pool. The present study uses an experimental framework in a single equity based artificial stock market as an electronic trading framework. This simulation could also be viewed as being too narrow to be representative of ETMs at large. However, this presents researchers with the opportunity to conduct additional research and work on additional simulations and also studies using field data to qualify present findings. The study also uses a limited (though statistically sufficient) number of participants for an experimental method which could be better supported by using a larger number of participants and repetition of the information conditions for more groups with multiple information artifacts for each TDIC condition. The present study used one information artifact, made available individually to participants, per TDIC condition, which has served as a strong starting point but creates a need for additional research to validate the present findings using additional artifacts. Past research has used student pools extensively and successfully and yet, in the present trading experiment, the absence of different types of traders, investors and experience levels in the participant pool limits the external validity of the experiment. However, this limitation by itself does not mitigate the finding that TDIC influences BTP in ETMs, though future research with a heterogeneous participant pool could further strengthen the present research.

# CONCLUSION

The results provide information systems, information science, electronic trading and behavioral finance researchers focused on technology and information management in electronic trading markets globally, with unique insights into the role of specific information categories in electronic markets. The findings provide further support for challenges associated by past research with decision making under the influence of varying information categories in electronic markets [25][35]. Specifically, the study shows the impact of behavioral aspects, including the effects of a strong belief in certain information categories such as "private information", Such beliefs can affect the trading performance of market participants who believe that they have superior information.

The findings of the present research point to a need to qualify two classical schools of thought involving first, the sufficiency of prices to reflect complete knowledge and secondly, a monolithic unitary perspective of information in markets. Belief in the sufficiency of prices in markets, based on an informal and unstructured aggregation of fragmented information possessed by participants, to lead to objective uniformity of prices will need to be qualified in light of the present finding that information categories can lead to behavioral effects which have the potential to distort informal price aggregation mechanisms in markets. Secondly, the assumption that information is unitary (information has no effects from variance in characteristics in any other dimension, except the dimension of objective content) is challenged by the finding of the present study which shows that there are significant performance effects even when objective information content (such as the expected stock price) is held constant, based on variations in information categories. This is an important finding for qualifying the understanding of information in many information-artifact based theories and models.

Our findings are aligned with recent information systems research which focus on multiple dimensions of information as seen in the incorporation of various information quality related characteristics in recent IS theories and models. Although bordering on conjecture, it is possible to extrapolate from the present research that information categories affect user beliefs, which shape expectations leading to decisions and finally use, which is cumulatively reflected as behavior, specifically BTP in our study as measured through the monetary trading performance measure – a potential topic for future research. The present research provides researchers and practitioners supporting behavioral theories in information science, finance and economics with additional information systems theory based corroboration of the presence of behavioral effects in ETMs. The present study also creates avenues for further research on the role of information systems in managing information categories and for exploring the role of TDIC in e-commerce. If belief in information category based behavioral effects are present in ETMs, it can have significant implications for practitioners, information management professionals, companies and consultants in ETMs and is suggestive of similar effects in e-commerce. Practitioners will need to factor in behavioral responses to information categories in decision support systems. Our findings can also influence information distribution and dissemination practices. Given the universality and commonalities of information characteristics areas across ETMs and e-commerce globally, it is hoped that the findings from this study can be extended with additional research, to be generalizable across domains.

# ACKNOWLEDGEMENTS

We would like to acknowledge the valuable support provided by Baruch College & the City University of New York in the form financial support through DSR Grant for funding the electronic trading experiment.

# AUTHOR BIOGRAPHIES

**Jim Samuel, Ph.D.,** is Associate Professor, School of Business, University of Charleston. His primary research covers human intelligence and artificial intelligences interaction, textual analytics, applied machine learning and socioeconomic implications of artificial intelligence. His research extends to innovation, women in advanced technologies, AI education and AI bias investigation. He holds M.Arch, M.B.A (International Finance) and Ph.D. (CIS) degrees (Updated, 2019).

**Richard D. Holowczak** is Associate Professor of CIS and is Director of the Wasserman Trading Floor / Subotnick Center at Baruch College, CUNY. He holds a BS/CS from the College of New Jersey, an MS/CS from the New Jersey Institute of Technology, and M.B.A. and Ph.D. degrees from Rutgers University.

**Alexander Pelaez** is an Assistant Professor of Information Technology and Business Analytics at Hofstra University. His research focuses on online purchases and group dynamics as well as the use of social networking and mobile technologies. His work has been published in the International Journal of Electronic Commerce, and Journal of Management Systems.